\documentclass[reqno]{amsart}
\usepackage{epsf}
\usepackage{anysize}
\marginsize{1in}{1in}{1in}{1.50in}

\begin{document}

\title{The Optimal Control Landscape for the Generation of Unitary Transformations with Constrained Dynamics}

\author{Michael Hsieh$^{1,2}$, Rebing Wu$^{5}$, Herschel Rabitz$^{6}$, Daniel Lidar$^{1,2,3,4}$ \\
\textit{$^{1}$Center for Quantum Information Science and Technology,
$^{2}$Department of Chemistry, $^{3}$Department of Electrical
Engineering, $^{4}$Department of Physics, University of Southern
California, Los Angeles, CA 90089, $^{5}$Department of Automation,
Tsinghua University, Beijing, 100084, P.R. China, $^{6}$Department
of Chemistry, Princeton University, Princeton, NJ 08544 }}
\date{\today}
\maketitle 

\begin{abstract}
The reliable and precise generation of quantum unitary
transformations is essential to the realization of a number of
fundamental objectives, such as quantum control and quantum
information processing. Prior work has explored the optimal control
problem of generating such unitary transformations as a surface
optimization problem over the quantum control landscape, defined as
a metric for realizing a desired unitary transformation as a
function of the control variables. It was found that under the
assumption of non-dissipative and controllable dynamics, the
landscape topology is trap-free, implying that any reasonable
optimization heuristic should be able to identify globally optimal
solutions. The present work is a control landscape analysis
incorporating specific constraints in the Hamiltonian corresponding
to certain dynamical symmetries in the underlying physical system.
It is found that the presence of such symmetries does not destroy
the trap-free topology. These findings expand the class of quantum
dynamical systems on which control problems are intrinsically
amenable to solution by optimal control.
\end{abstract}

\section{Introduction}
Central to many problems in quantum control \cite{BrumerShapiro} and
quantum information processing \cite{NielsenChuang} is the stable
and precise generation of specific unitary transformations. This
task may be viewed as an inverse problem where given a desired
unitary transformation, one must obtain the values of the control
variables of the system Hamiltonian whose dynamics generate it.

Generally, the connection between the control variables and the
unitary evolution operator is sufficiently complex such that the
exact control solution cannot be deduced from first principles. In
such cases adaptive optimization techniques are commonly applied,
such as optimal control theory (OCT) methods based on classical
variational optimization \cite{OptimalControlTechniques} for
computer simulations \cite{OCTexamples} and optimal control
experiment (OCE) methods based on evolutionary adaptation
\cite{JudsonRabitz} for practical laboratory studies
\cite{OCEexamples}. A method for the optimal generation of unitary
transformations based on OCT methods has been recently introduced
\cite{KosloffOCTUnitary}, with some promising successful
applications in simulation studies \cite{QuantumInfoExamples}.

A basic question is why such methods have had such a surprising
degree of success in such complex, high-dimensional problems. A step
toward the resolution of this question has been found in recent
theoretical studies of the \textit{quantum control landscape}
\cite{PRAunitaryPaper, UnitaryLandscapeFollowup}, defined generally
as the metric of attainment of some optimization objective as a
function of the control variables. In the present case, the
optimization objective is the generation of a specific unitary
transformation of the quantum system.

Consider a controllable quantum system defined on an $N$-level
Hilbert space whose dynamics are given by the time-dependent
Hamiltonian $H(t)$. The unitary evolution operator for the system is
\begin{equation}\label{E:TimeIntegration}
S(t_{i},t_{f}) = \mathbf{T}_{+} \thinspace \textup{exp} \left[
-\frac{i}{\hbar} \int_{t_{i}}^{t_{f}} dt \thinspace H(t) \right],
\end{equation}
where $\mathbf{T}_{+}$ denotes the time-ordering operation over some
time interval $[ t_{i}, t_{f} ]$. The optimization problem entails
choosing the controls within $H(t)$ that steer $S(t_{i},t_{f})$ to
some desired target transformation $W(t_{i}, t_{f}) \equiv W$. We
assume a degree of system controllability such that for any
$S(t_{i},t_{f}) \equiv S$ selected from the unitary Lie group
$U(N)$, there exists some choice of controls within $H(t)$ which
generates $S$ via Eq.(\ref{E:TimeIntegration}). This full access to
$U(N)$ makes it possible to define the control variables of the
problem to be some parametrization of $S$ itself, in lieu of the
controls within $H(t)$. This kinematic representation of the control
variables is particularly attractive in the sense that the analysis
now depends no longer on the specific structure of the Hamiltonian,
but only on the geometry of $U(N)$. Furthermore, the results are
generalizable to all finite-dimensional, controllable quantum
systems.

We employ as the distance measure between unitary transformations
the squared Hilbert-Schmidt metric $|| S - W ||^{2} = 2N - 2
\thinspace \textup{Re} \thinspace \textup{Tr} (W^{\dagger}S)$, whose
nonconstant part we define to be the \textit{landscape metric
function}
\begin{equation}\label{E:JHeightFunction}
J[S] = \textup{Re} \thinspace \textup{Tr} (W^{\dagger}S).
\end{equation}
Trace functionals of this type have been studied in a general
context \cite{VonNeumannPaper, FrankelBook}, as well as with
specific reference to physical \cite{GeissingerPaper} and control
theoretic applications \cite{ShaymanPaper, BrockettPaper}. We define
the \textit{control landscape} to be the image of the metric
function $J:U(N)\rightarrow \mathbb{R}$, where each point in $U(N)$
represents some choice of controls. We will refer to the space over
which the landscape is defined as the \textit{landscape domain},
which in this instance is $U(N)$.

The optimal control problem of generating unitary transformations of
a desired form is essentially a search over the control landscape
for critical regions at which the first-order variation of the
landscape function $J$ vanishes. In this sense, the difficulty of
the problem is largely determined by the topology of the critical
points. In a prior analysis of the problem of optimally generating
unitary transformations over $U(N)$ \cite{UnitaryLandscapeFollowup},
it was found that the number of disconnected critical regions of the
landscape corresponding to distinct values of $J$ scaled as $N+1$,
of which $N-1$ regions corresponded to local critical points with
saddlepoint topology, and the remaining to global extrema.
Furthermore, it was determined that the global extrema solutions
comprised zero-dimensional subspaces of $U(N)$, where the local
critical regions had the structure of complex Grassmannian
submanifolds embedded in $U(N)$ \cite{UnitaryLandscapeFollowup}.

Although the existence of such local (false) critical points may
still act as deleterious attractors for optimal searches, the number
of such regions grow only linearly with the Hilbert space dimension
of the underlying physical system. These appear to be competing
arguments in the assessment of whether this landscape is amenable to
adaptive searching methods. In recent numerical studies where
genetic algorithms, gradient following, and simplex methods have
been used, the observed exponential scaling of search convergence
time suggests that the deleterious influences may dominate
\cite{KWM_Paper}. Nevertheless, the desirable absence of any local
extremum traps over the entire class of such surfaces assures that
any optimization will eventually succeed in locating a global
solution.

A key open question is whether this desirable trap-free landscape
topology is preserved when the underlying dynamics of the physical
system are restricted or constrained in some manner. In the present
analysis, we consider various cases in which specific dynamical
symmetries are imposed on the system Hamiltonian, specifically
relating to total spin, space rotation invariance, and time-reversal
invariance. In prior analyses \cite{PRAunitaryPaper,
UnitaryLandscapeFollowup}, no structure of the Hamiltonian was
assumed other than its complex-valued Hermiticity (\textit{i.e.},
the imaginary part of the Hamiltonian is nonzero), which corresponds
to the symmetry class of systems without time-reversal invariance
\cite{DysonPaper}. We presently consider the topology of the control
landscape defined over physical systems with Hamiltonians from some
alternate symmetry classes:
\begin{enumerate}
  \item \noindent Systems with time-reversal invariance and integral total
  spin.
  \item \noindent Systems with time-reversal invariance and space rotation
symmetry.
  \item \noindent Systems with with time-reversal invariance, half-integer total spin, without
space rotation symmetry.
\end{enumerate}
The first two classes are described by \textit{real symmetric}
Hamiltonians  $H = H^{T}$. Such Hamiltonians arise in time-symmetric
pulsing strategies in spin control problems \cite{EBWBook,
Bonesteel_TRS_1,Bonesteel_TRS_2}, and continuous quantum random
walks \cite{Farhi_1,Farhi_2}. The third class is described by
\textit{symplectic} Hamiltonians $H = H^{R}$, where the $2N \times
2N$ matrix $H^{R}$ is the symplectic dual
\begin{equation}
H^{R} \equiv  -\left(
\begin{array}{ll} & \mathbb{I}_{N}  \\   -\mathbb{I}_{N} &
\end{array}
\right)H^{T}\left(
\begin{array}{ll} & \mathbb{I}_{N}  \\   -\mathbb{I}_{N} &
\end{array}
\right).
\end{equation}
Such Hamiltonians arise in the dynamics of Gaussian pure states
\cite{GPS_1}, which have notable applications in quantum optics
\cite{GPS_2}. A more extensive discussion connecting symmetry
classes of Hamiltonians with their corresponding physical systems
can be found in \cite{PorterBook, MehtaBook}

Section 2 presents formal definitions of the landscape function and
domain for the case of real symmetric Hamiltonians. From these
definitions, the remarkable property of the invariance of the
landscape topology with respect to the choice of target
transformation follows simply. The identification of the critical
landscape regions as a union of real Grassmannian submanifolds is
given in Section 3. The method for computing the signature of the
critical submanifolds is derived in Section 4. Section 5
recapitulates the analysis for the case of symplectic Hamiltonians.
Section 6 concludes.

\section{Landscape Function and Domain}

Consider the class of physical systems whose dynamics are described
by real symmetric Hamiltonians. Through
Eq.(\ref{E:TimeIntegration}), the dynamical propagators generated by
such Hamiltonians are symmetric unitary transformations. Let $S$ be
a programmable $N \times N$ symmetric unitary transformation
satisfying (i) $SS^{\dagger} = S^{\dagger}S = I_{N}$, where $I_{N}$
is the $N \times N$ identity, and (ii) $S = S^{T}$. These
constraints leave $\frac{N^{2}+N}{2}$ real degrees of freedom in $S$
\cite{MehtaBook}. For some desired symmetric unitary transformation
$W$, the optimization objective is to minimize the landscape metric
distance between $W$ and $S$ as defined in
Eq.(\ref{E:JHeightFunction}). In general, optimization heuristics
defined over the landscape will seek the \textit{critical points} of
the landscape where the first-order variation of the landscape
metric function $J$ vanishes. A remarkable quality of this
optimization problem is that the landscape topology is invariant to
the choice of the target transformation $W$. In this sense, the
optimization problems for all unitary transformations can be
expected to be of equivalent difficulty. We presently prove this
claim.

We assume that the range of controls is restricted such that the
Hamiltonian generating $S$ is always real symmetric. Any symmetric
unitary transformation $S$ has the canonical representation $S =
U^{T}U$ where $U \in U(N)$ \cite{MehtaBook}. Let $U_{1},U_{2} \in
U(N)$. The necessary and sufficient condition for $U_{1}^{T}U_{1} =
U_{2}^{T}U_{2}$ is that $U_{1}U_{2}^{-1} \in O(N)$ where $O(N)$
denotes the real orthogonal Lie group. Therefore, the image of the
canonical representation mapping $U \rightarrow U^{T}U$ is
homeomorphic to the homogeneous space $U(N) / O(N)$ \cite{Ktheory}.
Henceforth, we take the space $U(N) / O(N)$ of symmetric unitary
transformations to be the landscape domain.

Since $U(N)/O(N)$ is not closed under multiplication, we must
rearrange the argument of the trace function $J[S] = \textup{Re}
\thinspace \textup{Tr} \thinspace
(\sqrt{W^{\dagger}}S\sqrt{W^{\dagger}})$, where $S$ and $W$ are
symmetric unitary, to ensure that the landscape metric is defined
strictly over the landscape domain. Noting that
$\sqrt{W^{\dagger}}S\sqrt{W^{\dagger}} \rightarrow S$ is a
homeomorphism on $U(N)/O(N)$, the image of $J[\cdot] = \textup{Re}
\thinspace \textup{Tr} (\cdot)$ with the choice of $S$ as the
argument is equivalent to that with the choice of
$\sqrt{W^{\dagger}}S\sqrt{W^{\dagger}}$. Adopting the simpler choice
of $S$ as the argument, we obtain the topologically equivalent
landscape metric function
\begin{equation}
\mathcal{J}[S] = \textup{Re} \thinspace \textup{Tr} \thinspace (S)
\end{equation}
which has no dependence on the target transformation $W$. This
target-invariant landscape function $\mathcal{J}$ also has the
advantage of analytical simplicity, which we adopt for the remainder
of the analysis.

\section{Critical Submanifolds}

We presently determine the enumeration and topology of the landscape
regions on which $\mathcal{J}$ is critical. We will demonstrate that
the number of such regions scales linearly with $N$ and has the
structure of real Grassmannians embedded in $U(N)/O(N)$.

Any symmetric unitary transformation can be diagonalized as
\begin{equation}\label{E:Rotational_Representation}
S = X^{T} \Omega X
\end{equation}
where $X$ is an element of the real special orthogonal Lie group
$SO(N)$ and $\Omega$ is a diagonal operator
\begin{equation}
\Omega = \left(
\begin{array}{lll} e^{i\varphi_{1}} & & \\ & \ddots & \\ & & e^{i\varphi_{N}}
\end{array}
\right).
\end{equation}
of the unimodular eigenvalues
$\{e^{i\varphi_{1}},...,e^{i\varphi_{N}}\}$ of $S$ \cite{HuaBook}.
Using the cyclic property of the trace, the target-invariant
landscape metric function simplifies to
\begin{align}
\mathcal{J}[S] &= \textup{Re} \thinspace \textup{Tr} (\Omega) \label{E:NewDefinitionForJ} \\
&= \sum_{j=1}^{N} \textup{cos} \thinspace \varphi_{j}.
\label{E:CostFunctionInAngles}
\end{align}
From Eq.(\ref{E:CostFunctionInAngles}), we see that the first-order
variation $\delta \mathcal{J} = -\sum_{j=1}^{N} \textup{sin}
\thinspace \varphi_{j} \thinspace d\varphi_{j}$ vanishes when
$\varphi_{j} = \ell_{j} \pi$, for any integers $\ell_{j}$. There are
$N+1$ critical values $-N,-N+2,...,N-2,N$ for $\mathcal{J}$,
determined by the parities of $\ell_{j}, j=1,...,N$. The
corresponding critical points $\tilde{S} = X^{T}\tilde{\Omega}^{(n)}
X$ comprise equivalence classes of transformations orthogonally
similar to canonical elements $\tilde{\Omega}^{(n)}$ where $n$
denotes the number of even integers in the set $\{
\ell_{1},...,\ell_{N} \}$:
\begin{equation}
\tilde{\Omega}^{(n)} = \left(
\begin{array}{ll} \mathbb{I}_{n} & \\  & -\mathbb{I}_{N-n}
\end{array}
\right).
\end{equation}
We seek to determine the topology of the $N+1$ equivalence classes
of transformations corresponding to each distinct critical value.
Such equivalence classes are the \textit{critical submanifolds} of
the landscape, composed of points at which the first-order
optimization condition $\nabla \mathcal{J} = 0$ is satisfied.

Consider the conjugation of $\tilde{\Omega}^{(n)} \in U(N)/O(N)$ by
some $\Xi \in SO(N)$ as a group action $\mathcal{G}: SO(N) \times
U(N)/O(N) \rightarrow U(N)/O(N)$
\begin{equation}
\mathcal{G} \cdot \tilde{\Omega}^{(n)} = \Xi^{T}
\tilde{\Omega}^{(n)} \Xi,
\end{equation}
for which $SO(N)$ is the acting group and $U(N)/O(N)$ is the
$\mathcal{G}$-space. The stabilizer (isotropy) subgroup
$\textup{STAB}(\tilde{\Omega}^{(n)})$ of any $\tilde{\Omega}^{(n)}$,
defined as the subgroup of the acting group $SO(N)$ whose elements
map $\tilde{\Omega}^{(n)}$ back to itself via the
$\mathcal{G}$-action, is composed exclusively of elements of the
form
\begin{equation}
\tilde{\Xi} = \left(
\begin{array}{ll} \Xi_{n} & \\  & \Xi_{N-n}
\end{array}
\right)
\end{equation}
where $\Xi_{n} \in SO(n)$ and $\Xi_{N-n} \in SO(N-n)$. Therefore,
the stabilizer is a product subgroup of $SO(N)$:
\begin{equation}
\textup{STAB}(\tilde{\Omega}^{(n)}) = SO(n) \times SO(N-n).
\end{equation}
Define a mapping which associates a fixed $\mathcal{G}$-space
element $\Omega $ with \textit{some} element of the acting group:
\begin{equation}
\mathcal{G}_{\Omega}: SO(N) \rightarrow U(N)/O(N), \quad \Omega
\rightarrow  \mathcal{G} \cdot \Omega.
\end{equation}
When the domain of $\mathcal{G}_{\Omega}$ is taken to be all of
$SO(N)$, the image is simply the \textit{orbit} of $\Omega$:
\begin{equation}
\mathcal{O}(\Omega ) \equiv \{ \Xi^{T} \Omega \Xi : \Xi \in SO(N)\},
\end{equation}
These orbits, for $\Omega = \tilde{\Omega}^{(n)}$, where
$n=0,...,N$, comprise precisely the $N+1$ sets of critical points of
the landscape.

To obtain their topological structure, we recall that by the
orbit-stabilizer theorem, $\mathcal{G}_{\Omega}$ induces a bijection
$SO(N) / \textup{STAB}(\tilde{\Omega}^{(n)}) \rightarrow
\mathcal{O}(\tilde{\Omega}^{(n)})$, which we may sharpen further to
be a diffeomorphism because $SO(N)$ is a compact Lie group
\cite{HelmkeMooreBook}. Therefore, the critical set is composed of
the union
\begin{equation}
\bigcup_{n=0}^{N} G_{real}(n,N)
\end{equation}
where
\begin{equation}
G_{real}(n,N) = SO(N) / SO(n) \times SO(N-n)
\end{equation}
embedded in $U(N)/O(N)$. The dimensionality of the Grassmannian is
well known to be
\begin{align}\label{E:CM_Dimensionality}
\textup{dim} \thinspace G_{real}(n,N) &= \frac{N^{2} - N}{2} -
\left[ \frac{n^{2} - n}{2}  + \frac{(N-n)^{2} - N + n}{2} \right] \\
&= n(N-n).
\end{align}
Since the objective of any optimization is to attain the
$\mathcal{J} = N$ or $-N$ values, corresponding to a perfect
generation of the desired transformation, an immediate consequence
of the foregoing dimensionality equation is that the critical
submanifolds corresponding to non-global critical have nonzero
dimension in appropriate subspaces of $U(N)/O(N)$ whereas the
critical submanifolds corresponding to global critical points, with
$n = 0$ or $n=N$, strictly have dimension zero. It is therefore of
practical importance to determine whether the critical submanifolds
corresponding to non-global critical points have the topology of
local maxima or minima, which may act as traps, or of non-trapping
saddlepoints.

\section{Hessian Analysis of Critical Points}
A critical point of $\mathcal{J}$ can be identified as an extremum
or a saddlepoint by computing its signature $\left(
D_{+},D_{-},D_{0} \right)$, an ordered triple of integers denoting
the number of upward, downward and flat landscape directions at that
point. $D_{+}$ and $D_{-}$ are commonly referred to as the indices
of positive and negative inertia, and $D_{0}$ as the kernel
dimension. The Hessian operator of $\mathcal{J}$ evaluated at
$\tilde{S}$ is defined
\begin{equation}
\mathcal{H}_{i^{\prime}i} =
\frac{d^{2}\mathcal{J[\tilde{S}]}}{dx_{i^{\prime}} \thinspace
dx_{i}},
\end{equation}
where $\{ x_{i} \}_{i=1,...,\frac{N^{2} + N}{2}}$ is some set of
local coordinates around a point $S \in U(N)/O(N)$. Since the
Hessian is symmetric, there exists some coordinate transformations
$x_{i} \rightarrow \hat{x}_{i}$ which rotates $\mathcal{H}$ into
diagonal form, where the enumeration of its positive, negative and
zero-valued eigenvalues corresponds exactly to $D_{+}, D_{-}$ and
$D_{0}$. By Sylvester's Law of Inertia, $D_{+}, D_{-}$ and $D_{0}$
are invariant to changes in the coordinate system
\cite{HornJohnsonBook}.

As a symmetric matrix, $\mathcal{H}$ is representable as a quadratic
form $\langle \Gamma | Q(\mathcal{H}) | \Gamma \rangle = \sum_{i,j}
Q_{i,j}\Gamma_{i}\Gamma_{j}$, with real coefficients $\{ Q_{i,j} \}$
and $\{ \Gamma_{i}, \Gamma_{j} \}$ denoting components of a vector
$| \Gamma \rangle$ of real indeterminates. A corollary of
Sylvester's Law assures that the coordinate rotation $x_{i}
\rightarrow \hat{x}_{i}$ induces a transformation of the quadratic
form into a canonical form of strictly second degree monomials
\begin{equation}\label{E:General_Quadratic_Form}
\langle \Gamma | \hat{Q}(\mathcal{H}) | \Gamma \rangle = \sum_{i}
\hat{Q}_{i}\hat{\Gamma}_{i}^{2},
\end{equation}
where the enumeration of the positive, negative and zero-valued real
coefficients $\hat{Q}_{j}$ corresponds to $D_{+}, D_{-}$ and $D_{0}$
\cite{HornJohnsonBook}.

To explicitly connect the Hessian matrix with its quadratic form,
let us consider $\mathcal{J}$ as a mapping $\left( \mathcal{J} \circ
\gamma \right)(t)$ over a parameterized arc $\gamma (t) \in
U(N)/O(N)$ satisfying $\gamma (0) = \tilde{S}$. Taking the second
derivative of the arc-parameterized $\mathcal{J}$ at $t=0$, we have
\cite{DemazureBook}
\begin{align}
\left( \mathcal{J} \circ \gamma \right)^{\prime \prime}(0) &= \left(
\sum_{i} \frac{\partial \mathcal{J} \left[ \gamma (0) \right]
}{\partial
x_{i}} \dot{\gamma}_{i} (0) \right)^{\prime} \\
&= \sum_{i} \frac{\partial \mathcal{J} \left[ \gamma (0) \right]
}{\partial x_{i}} \ddot{\gamma}_{i}(0) + \sum_{i,j}
\frac{\partial^{2} \mathcal{J} \left[ \gamma (0) \right]}{\partial
x_{i}\partial x_{j}} \dot{\gamma}_{i}(0) \dot{\gamma}_{j}(0).
\end{align}
Noting that $\frac{\partial \mathcal{J} \left[ \gamma (0) \right]
}{\partial x_{i}}$ vanishes since $\tilde{S}$ is critical, we
identify the remaining term $\sum_{i,j} \frac{\partial^{2}
\mathcal{J} \left[ \gamma (0) \right]}{\partial x_{i}\partial x_{j}}
\dot{\gamma}_{i}(0) \dot{\gamma}_{j}(0)$ as a quadratic form mapping
tangent space vectors $\dot{\gamma}(0) \in T_{\tilde{S}} \thinspace
U(N)/O(N)$ to the reals \cite{DemazureBook}. Transforming into the
diagonal coordinate system $\{ \hat{x}_{i} \}$, we have
\begin{equation}
\left( \mathcal{J} \circ \gamma \right)^{\prime \prime}(t) =
\sum_{i} \frac{\partial^{2} \mathcal{J} \left[ \gamma (t)
\right]}{\partial \hat{x}_{i}^{2}} \dot{\gamma}_{i}^{2}(0).
\end{equation}
Associating $\frac{\partial^{2} \mathcal{J} \left[ \gamma (t)
\right]}{\partial \hat{x}_{i}^{2}} \leftrightarrow \hat{Q}_{i}$ and
$\gamma_{i}^{2}(0) \leftrightarrow \hat{\Gamma}_{i}^{2}$, we obtain
a direct identification of the Hessian matrix with its quadratic
form.

We now compute the Hessian quadratic form (HQF) explicitly. Let us
evaluate $\mathcal{J}$ at some critical point $\tilde{S}$ in the
previously defined representation $\tilde{S} = X^{T}\tilde{\Omega}
X$. Since $\tilde{S}$ is a critical point of $\mathcal{J}$,
$\tilde{\Omega}$ takes the form
\begin{equation}
\tilde{\Omega} = \left(
\begin{array}{lll} \tilde{\omega}_{1} & & \\ & \ddots & \\ & & \tilde{\omega}_{N}
\end{array}
\right).
\end{equation}
where $\tilde{\omega}_{1} = \pm 1,...,\tilde{\omega}_{N} = \pm 1$.
If $\tilde{\omega}_{j} = +1 (-1)$ for all $j$, we are at a global
maximum (minimum). Otherwise, we are at a local critical point. We
presently seek to establish that these critical points are not local
extremum traps.

To obtain the HQF, we twice differentiate the landscape function
argument $S$ along the parameterized curve
$S=\sqrt{\tilde{S}}e^{iAt}\sqrt{\tilde{S}}$ in $U(N)/O(N)$ defined
by some real matrix of indeterminates $A = A^{T}$:
\begin{equation}
S \rightarrow S^{\prime} = \frac{d}{dt}\Big|_{t=0}
\sqrt{\tilde{S}}e^{iAt}\sqrt{\tilde{S}} =
\sqrt{\tilde{S}}{iA}\sqrt{\tilde{S}} \rightarrow S^{\prime \prime} =
- \sqrt{\tilde{S}}A^{2}\sqrt{\tilde{S}}
\end{equation}
and evaluate $\mathcal{J}$ explicitly in terms of the matrix
elements of $A$.
\begin{align}
\mathcal{H}_{\mathcal{J}}(S) &\equiv -\textup{Re} \thinspace
\textup{Tr}
\thinspace \left(A^{2}S\right) \\
&= - \textup{Re} \thinspace \textup{Tr}
\thinspace \left(X A^{2}X^T\tilde\Omega \right) \\
&= - \textup{Re} \thinspace \textup{Tr}
\thinspace \left(\tilde{A}^{2}\tilde\Omega\right) \\
&= -\sum_{i \neq j}\tilde{\omega}_{i} \tilde{A}_{ij}^{2} -
\sum_{i}\tilde{\omega}_{i}\tilde{A}_{ii}^{2} \\
&= -\sum_{i < j}(\tilde{\omega}_{i} +
\tilde{\omega}_{j})\tilde{A}_{ij}^{2} -
\sum_{i}\tilde{\omega}_{i}\tilde{A}_{ii}^{2},\label{E:HQF_Final_Formula}
\end{align}
where $\tilde{A} \equiv X A X^T$ and multiplicative constants have
been normalized to unity.

To determine the values of $D_{+},D_{-},D_{0}$, we note that for a
critical point $\tilde{S}$ with eigenvalues $+1$ and $-1$ with
multiplicities $n$ and $N-n$, the only nonzero coefficients in the
first summation $-\sum_{i < j}(\tilde{\omega}_{i} +
\tilde{\omega}_{j})\tilde{A}_{ij}^{2}$ of
Eq.(\ref{E:HQF_Final_Formula}) will be generated by
$\tilde{\omega}_{i} = \tilde{\omega}_{j} = +1$ or
$\tilde{\omega}_{i} = \tilde{\omega}_{j} = -1$. There are
$\frac{n(n-1)}{2}$ ways of selecting indices $i$ and $j$ satisfying
$i < j$ from the $n$-fold set of indices corresponding to positive
values $\{ k \thinspace : \thinspace \tilde{\omega}_{k} = +1 \}$,
and thus $\frac{n(n-1)}{2}$ negative-valued monomials
$\tilde{A}_{ij}^{2}$ in this summation. We add this to the $n$
additional negative monomials from the second summation
$-\sum_{i}\tilde{\omega}_{i}\tilde{A}_{ii}^{2}$ to obtain $D_{-} =
\frac{n^{2} + n}{2}$. By similar counting, it can be seen that there
are $D_{+} = \frac{(N-n)^{2} + N-n}{2}$ positive monomials. Noting
that the sum of the indices of inertia and the kernel dimension must
equal the dimensionality of the domain of the function whose Hessian
is being computed
\begin{align}
D_{+}+D_{-}+D_{0} &= \textup{dim} \thinspace U(N)/O(N) \\
&= \frac{N^{2} + N}{2},
\end{align}
it is evident that
\begin{align}
D_{+} &= \frac{(N-n)^{2} + N-n}{2}\\
D_{-} &= \frac{n^{2} + n}{2}\\
D_{0} &= n(N-n).
\end{align}
Finally, it is evident that for any local critical point for which
both $N,N-n > 0$, both $D_{+}$ and $D_{-}$ must be nonzero.
Therefore, all local critical points are assured to possess
saddlepoint topology, and cannot act as local extremum traps.

\section{Control Landscape Topology for Symplectic Hamiltonians}

We consider the class of physical systems whose dynamics are
described by symplectic Hamiltonians. Since the analysis is almost
identical to the real symmetric Hamiltonian case, the discussion
will be limited to statements of the key assumptions and
conclusions. Let $S$ denote programmable symplectic unitary
dynamical operator generated by such a Hamiltonian via
Eq.(\ref{E:TimeIntegration}) with quaternion-real matrix elements in
the $2 \times 2$ Pauli basis $\{
\sigma_{0},\sigma_{X},\sigma_{Y},\sigma_{Z} \}$. The choice of the
Pauli basis for the matrix elements induces a $2N \times 2N$
representation of $S$, which must satisfy (i) $SS^{\dagger} =
S^{\dagger}S = I_{2N}$ and (ii) $S = S^{R}$ where $S^{R}$ denotes
the symplectic dual. Such constraints leave $2N(N-1)$ real degrees
of freedom for $S$.

In analogy to the prior case, we assume that the range of controls
is restricted such that the Hamiltonian generating $S$ is always
symplectic. Any symplectic dual transformation $S$ has the canonical
representation $S = U^{R}U$ where $U \in U(2N)$ \cite{MehtaBook}.
Let $U_{1},U_{2} \in U(2N)$. The necessary and sufficient condition
for $U_{1}^{T}U_{1} = U_{2}^{T}U_{2}$ is that $U_{1}U_{2}^{-1} \in
Sp(2N)$ where $Sp(2N)$ denotes the real symplectic Lie group. The
image of the mapping giving the canonical representation $U
\rightarrow U^{R}U$ is homeomorphic to $U(2N) / Sp(2N)$. Henceforth,
we take $U(2N) / Sp(2N)$ to be the space of symplectic unitary
transformations over which we define as the landscape domain for the
present case. In close analogy with the prior case, we may also
define a target-invariant landscape function $\mathcal{J}[S] =
\textup{Re} \thinspace \textup{Tr} \thinspace (S)$ defined over
$U(2N) / Sp(2N)$.

The first and second order analyses of the critical landscape
topology can be replicated for the symplectic case by using, in
analogy to Eq.(\ref{E:Rotational_Representation}), the canonical
representation for unitary symplectic operators
\begin{equation}
S = X^{R}\Omega X,
\end{equation}
with $X \in Sp(2N)$ and
\begin{equation}
\Omega = \left(
\begin{array}{lll} \omega_{1} \sigma_{0}& & \\ & \ddots & \\ & & \omega_{N}\sigma_{0}
\end{array}
\right),
\end{equation}
where $\omega_{1},...,\omega_{N}$ are the complex, unimodular
eigenvalues of $S$.

Analogously to the prior case, there are $N+1$ critical submanifolds
with the structure of symplectic Grassmannians. Collectively, the
critical set is the union
\begin{equation}
\bigcup_{n=0}^{N} G_{symplectic}(2n, 2N)
\end{equation}
where
\begin{equation}
G_{symplectic}(2n, 2N) = Sp(2N) / Sp(2N-2n) \times Sp(2n)
\end{equation}
embedded in $U(2N) / Sp(2N)$, with dimension
\begin{equation}
\textup{dim} \thinspace G_{symplectic}(2n, 2N) = N(2N-1) -
(N-n)[2(N-n)-1] - n[2n - 1].
\end{equation}

The HQF can be obtained similarly. Let us evaluate $\mathcal{J}$ at
some critical point $\tilde{S}$ in the representation $\tilde{S} =
X^{R}\tilde{\Omega} X$. In analogy to the prior case, we
differentiate $\tilde{S}$ twice over the arc given by $\tilde{S}
\rightarrow S^{\prime} = \frac{d \tilde{S}}{dt}\Big|_{t=0}=
\sqrt{\tilde{S}} iA \sqrt{\tilde{S}} \rightarrow S^{\prime \prime} =
-\sqrt{\tilde{S}} A^{2} \sqrt{\tilde{S}}$ defined by some $A =
A^{R}$ (where $A_{ij} = \alpha_{ij}\sigma_{0} + \beta_{ij}\sigma_{x}
+ \gamma_{ij}\sigma_{y} + \delta_{ij}\sigma_{z}$ with real
$\alpha_{ij},\beta_{ij},\gamma_{ij},\delta_{ij}$) and evaluate the
landscape function with the arc-parameterized argument to obtain, up
to multiplicative constants
\begin{equation}
\mathcal{H}_{\mathcal{J}}(S) = -\sum_{i < j}(\omega_{i} +
\omega_{j})(\alpha_{ij}^{2}+\beta_{ij}^{2}+\delta_{ij}^{2}+\gamma_{ij}^{2})
- \sum_{i}\omega_{i}\lambda_{i}^{2}.
\end{equation}
The number of positive, negative and zero-valued elements of the HQF
are, respectively,
\begin{align}
D_{+} &= 2(N-n)^{2} + 2(N-n)\\
D_{-} &= 2n^{2} + 2n\\
D_{0} &= 4N(N-n).
\end{align}

\section{Conclusions}

The analysis reveals that the critical submanifolds for the
symmetry-restricted landscapes are of two types: isolated points
corresponding to the global maxima and minima, and Grassmannian
submanifolds corresponding to the sub-optimal extrema values.
Although the sub-optimal Grassmannian solutions are more numerous
and more voluminous than the global solutions, the Hessian analysis
reveals that all such local solutions have saddlepoint structure and
thus do not act as local traps. Furthermore, the invariance of the
qualitative landscape structure with respect to the target
transformation demonstrated in this analysis suggests that no
quantum unitary transformation should be any more difficult to
optimally construct than another. In addition to prior works
\cite{PRAunitaryPaper, UnitaryLandscapeFollowup}, the present
analysis is a strong endorsement of optimal search as a practically
viable means of generating unitary transformations for complex
systems, and demonstrates that the most important qualitative
aspects of the landscape are not affected when certain symmetry
restrictions are placed on the underlying dynamics.

\end{document}